\documentclass[12pt,a4paper]{article}
\usepackage{a4,epsf,here,epsfig}
\newcommand{\bmath}{\begin{eqnarray}}
\newcommand{\emath}{\end{eqnarray}}

\newcommand{\sg}{\mbox{sgn}}
\newcommand{\ab}{a_B}
\newcommand{\ua}{\uparrow}
\newcommand{\da}{\downarrow}
\newcommand{\si}{\sigma}
\newcommand{\kk}{\kappa}

\newcommand{\eqref}[1]{{{(}\nolinebreak\ref{#1})\nolinebreak}}

\begin{document}

\noindent
{\large\bf  Spin-resolved second-order correlation energy\\
of the two-dimensional uniform electron gas}

\medskip

\noindent
Michael Seidl, Institute of Theoretical Physics, University of Regensburg,\\
D-93040 Regensburg, Germany

\bigskip

\noindent
{\small
{\bf Abstract.} For the two-dimensional electron gas, the exact high-density limit
of the correlation energy is evaluated here numerically for all values of the spin
polarization. The result is spin-resolved into
$\uparrow\uparrow$, $\uparrow\downarrow$, and $\downarrow\downarrow$
contributions and parametrized analytically. Interaction-strength interpolation
yields a simple model (LSD) for the correlation energy at finite densities.
}

\medskip

\noindent
In recent years, two-dimensional (2D) electron systems have become the subject of
extensive research \cite{AKS}. The 2D version of density functional theory (DFT)
has proven particularly successful in studying quantum dots \cite{Steffi,JBY,Saarikoski}.
The local spin-density approximation (LSD) of DFT requires the correlation energy
of the spin-polarized uniform electron gas. This quantity in 2D is known accurately
for a wide range of densities and spin polarizations from
fixed-node diffusion Monte Carlo simulations \cite{AMGB}. Its high-density limit
is known exactly in terms of six-dimensional momentum-space integrals \cite{RK}.
Resolved into contributions due to $\ua\ua$, $\ua\da$, and $\da\da$ excitation
electron pairs, these integrals are evaluated here numerically. The analytical
parametrization of the results, Eqs.~\eqref{result1} and \eqref{result2} below,
is a crucial ingredient for the construction of the spin-resolved correlation energy at
finite densities, performed recently for the 3D electron gas \cite{PaolaJohn}.
It is also required for studying the magnetic response of the spin-polarized 2D
electron gas \cite{MM,PT}.
Generally, it provides a fundamental test for numerical parametrizations of the
correlation energy \cite{AMGB}.

In the 2D uniform electron gas, the electrons are moving on a plane at uniform
density $\rho\!=\![\pi(r_s\ab)^2]^{-1}$, where $\ab\!=\!0.529$ \AA ~is the Bohr radius
and $r_s$ is the dimensionless density parameter (Seitz radius). 
We consider lowest-energy states with a given spin polarization
\bmath
\zeta\;\equiv\;\frac{\rho_\uparrow-\rho_\downarrow}{\rho}
\emath
where $\rho_\uparrow$ and $\rho_\downarrow\equiv\rho-\rho_\uparrow$, respectively, are the (uniform)
densities of spin-up and spin-down electrons.
Including a neutralizing positive background, the total energy per electron is a unique
function of the dimensionless parameters $r_s$ and $\zeta$,
\bmath
e_{tot}(r_s,\zeta)\;=\;t_s(r_s,\zeta ) + e_x(r_s,\zeta ) + e_c(r_s,\zeta ).     
\label{etot}
\emath
The non-interacting kinetic and exchange energies,
\bmath
t_s(r_s,\zeta) \;=\; \frac{1+\zeta^2}2\,\frac1{r_s^2}\,,\qquad\qquad
e_x(r_s,\zeta) \;=\; -\frac{4\sqrt{2}}{3\pi}\,
        \frac{(1\!+\!\zeta )^{3/2}+(1\!-\!\zeta )^{3/2}}2
             \,\frac1{r_s}
\label{tsex}
\emath
(all energies are given in units of 1 Ha $\equiv e^2/a_B=$ 27.21 eV in the following),
may be understood as the 0th- and the 1st-order terms of a perturbation expansion for
the electron-electron interaction (where $r_s$ turns out to be the expansion parameter).

The remaining {\em correlation energy} in Eq.~\eqref{etot}
appears to have the perturbation (high-density) expansion \cite{Macke,CM}
\bmath
e_c(r_s,\zeta)\;=\;\sum\limits_{n=0}^{\infty}
\Big[a_n(\zeta)\,\ln(r_s)\,+\,b_n(\zeta)\Big]\,r_s^n \qquad\qquad (r_s\ll1).
\label{ecHD}
\emath
For the 2D electron gas (but not for the 3D one), the first coefficient vanishes,
$a_0(\zeta)\equiv0$. Consequently, the second-order ($n=0$) term is
$e_c^{(2)}(\zeta)\equiv b_0(\zeta)$, representing the high-density ($r_s\to0$) limit
of $e_c(r_s,\zeta)$. It can be split into an exchange (``$2b$'') and a ring-diagram
(``$2r$'') term \cite{RK},
\bmath
e_c^{(2)}(\zeta)\;=\;e_c^{(2b)}+e_c^{(2r)}(\zeta).
\emath

The exchange term has only equal-spins contributions,
$e_c^{(2b)}\!=\!e_{c\ua\ua}^{(2b)}(\zeta)\!+\!e_{c\da\da}^{(2b)}(\zeta)$,
given by the $\delta_{\si_1\si_2}$ term of Eq.~(14) in Ref.~\cite{RK}
(we choose the $k_x$ axis in the direction of ${\bf q}$),
\bmath
e_{c,\si\si}^{(2b)}(\zeta)\;=\;\frac1{8\pi^2}\int_0^{\infty}\frac{dq}q
\int\limits_{A[\kk_\si(\zeta),q]}\!\!\!\!\!\!d^2k_1
\int\limits_{A[\kk_\si(\zeta),q]}\!\!\!\!\!\!d^2k_2\;
\frac1{|q\,{\bf e}_x+{\bf k}_1+{\bf k}_2|}\;\frac1{q+k_{1x}+k_{2x}}.
\label{ec2b}
\emath
Here, $q$, ${\bf k}_1$, and ${\bf k}_2$ are dimensionless,
$\si\in\{\ua,\da\}$, and the domain of the 2D integrals is
\bmath
A[\kk,q]\;\equiv\;
\Bigl\{\;{\bf k}\in{\sf R}^2\;\Bigl|\;|{\bf k}|<\kk,\;\;|{\bf k}+q\,{\bf e}_x|>\kk\Bigl\},\qquad\quad
\kk_\si(\zeta)\equiv\Big[1+\sg(\si)\zeta\Big]^{1/2}.
\label{domain}
\emath
[$\kk_\si(\zeta)$ is the Fermi wave vector for spin-$\si$ electrons in units of its value at $\zeta=0$.]
Scaling the integration variables by some constant $\kk$, $q=\kk Q$ and ${\bf k}=\kk{\bf K}$, we have
generally
\bmath
\int_0^{\infty}\frac{dq}q\int\limits_{A[\kk,q]}d^2k\,f(q,{\bf k})
\;=\;\kk^2\int_0^{\infty}\frac{dQ}Q\int\limits_{A[1,Q]}d^2K\,f(\kk q,\kk{\bf K}).
\emath
Applying this rule to the integrals in Eq.~\eqref{ec2b}, we find \cite{RK}
\bmath
e_{c,\si\si}^{(2b)}(\zeta)\;=\;\Big[1+\sg(\si)\zeta\Big]\,J^{(2b)}.
\emath
Consequently \cite{RK}, the full second-order exchange term
$e_c^{(2b)}=e_{c\ua\ua}^{(2b)}(\zeta)+e_{c\da\da}^{(2b)}(\zeta)\equiv 2J^{(2b)}$
is $\zeta$-independent. A Monte Carlo integration yields
\bmath
J^{(2b)}\;\equiv\;e_{c\ua\ua}^{(2b)}(0)\;=\;(57.15\pm0.05)\,\mbox{mHa}\qquad\qquad
(1\mbox{mHa}=10^{-3}\mbox{Ha}).
\label{Jb}
\emath

\newpage
The ring-diagram term $e_c^{(2r)}(\zeta)$ is the remaining part of expression (14)
in Ref.~\cite{RK}, with the contributions
\bmath
e_{c,\si_1\si_2}^{(2r)}(\zeta)\;=\;
-\frac1{8\pi^2}\int_0^{\infty}\frac{dq}{q^2}\,
\int\limits _{A[\kk_{\sigma_1}(\zeta),q]}\!\!\!\!\!\!d^2k_1\;
\int\limits _{A[\kk_{\sigma_2}(\zeta),q]}\!\!\!\!\!\!d^2k_2\;\frac1{q + k_{1x}+k_{2x}}.
\label{intMC}
\emath 
The equal-spins terms ($\si_1=\si_2$) can be treated in the same way as the integral \eqref{ec2b},
\bmath
e_{c,\si\si}^{(2r)}(\zeta)\;=\;-\Big[1+\sg(\si)\zeta\Big]\,J^{(2r)},\qquad\qquad
J^{(2r)}\;=\;(76.69\pm0.03)\,\mbox{mHa}.
\label{Jr}
\emath
The only non-trivial $\zeta$-dependence is in the opposite-spins term
$e_{c\ua\da}^{(2r)}(\zeta)\equiv e_{c\da\ua}^{(2r)}(\zeta)$,
\bmath
e_{c\ua\da}^{(2r)}(\zeta)\;=\;e_{c\ua\da}^{(2r)}(0)\,\Big[1\!-\!f(\zeta)\Big].
\label{fdef}
\emath
By definition, $f(0)=0$, and, since $A[\kk_{\da}(1),q]=\emptyset$, $f(1)=1$. Moreover,
$e_{c\ua\da}^{(2r)}(0)=-J^{(2r)}$. When the results of a Monte Carlo evaluation of
$f(\zeta)$ at different values of $\zeta$ are compared with the functions
$f_\alpha(\zeta)\equiv[(1\!+\!\zeta)^\alpha+(1\!-\!\zeta)^\alpha-2]/(2^\alpha-2)$,
particularly good agreement (specially for $\zeta\to0$ and $\zeta\to1$) is
found in the limit $\alpha\to1$ (Fig.~1a),
\bmath
f(\zeta)\;=\;f_1(\zeta)+\delta f(\zeta),\qquad\qquad
f_1(\zeta)
\;\equiv\;\frac{(1\!+\!\zeta)\ln(1\!+\!\zeta)\,+\,(1\!-\!\zeta)\ln(1\!-\!\zeta)}{2\ln2}.
\label{fzeta}
\emath
[Note that $f_\alpha(\zeta)$ also represents the $\zeta$-dependence of $t_s$
($\alpha=2$) and $e_x$ ($\alpha=\frac32$) in Eq.~\eqref{tsex}.] The small deviation
$\delta f(\zeta)$ is accurately fitted by a polynomial (Fig.~1b)
\bmath
\delta f(\zeta)\;\approx\;0.0636\,\zeta^2\,-\,0.1024\,\zeta^4\,+\,0.0389\,\zeta^6\,.
\label{dfzeta}
\emath
The small minimum of $\delta f(\zeta)$ indicated by the numerical data (dots in Fig.~1b)
at $\zeta\approx0.98$ is probably real, since a similar peculiarity is observed for the 3D
electron gas (see the inset in Fig.~1 of Ref.~\cite{Hoffman}).

In summary, the second-order correlation energy $e_c^{(2)}(\zeta)=e_c^{(2b)}+e_c^{(2r)}(\zeta)$ is
\bmath
e_c^{(2)}(\zeta)\;\equiv\;e_{c\ua\ua}^{(2)}(\zeta)+2e_{c\ua\da}^{(2)}(\zeta)+e_{c\da\da}^{(2)}(\zeta)
\;=\;\Big[153.38\,f(\zeta)-192.46\Big]\,\mbox{mHa},
\label{result1}
\emath
where $f(\zeta)$ is given by Eqs.~\eqref{fzeta} and \eqref{dfzeta}.
The spin resolution is fixed by
\bmath
e_{c\ua\ua}^{(2)}(\zeta)\;\equiv\;e_{c\da\da}^{(2)}(-\zeta)\;=\;-(1+\zeta)\times19.54\,\mbox{mHa}.
\label{result2}
\emath

$e_c^{(2)}(\zeta)\equiv e_c(0,\zeta)$ is the high-density limit of the general correlation energy
$e_c(r_s,\zeta)$. 
To illustrate the relevance of this limit for finite densities ($r_s>0$), the present result
can be used in the interaction-strength interpolation (ISI) of Ref.~\cite{ISI}. This approach
does not require the higher-order ($n\ge1$) terms of the expansion \eqref{ecHD} (which is
expected to have only a finite radius of convergence).
Instead, information beyond the second order is taken from the low-density (strong-interaction
or Wigner-crystal) limit of the exchange-correlation energy $e_{xc}\equiv e_x+e_c$ (per electron),
\bmath
e_{xc}(r_s,\zeta)
\;\to\;\frac{a_{\infty}}{r_s}+\frac{b_{\infty}}{r_s^{3/2}}\qquad\qquad(r_s\to\infty).
\emath
The coefficients \cite{WigCry} $a_{\infty}\approx-1.1061$ and $b_{\infty}\approx\frac12$
are independent of $\zeta$, since any spatial overlap between two electrons is strongly
suppressed in this limit, no matter whether their spins are parallel or not \cite{PC}. The
resulting ISI expression for the exchange-correlation energy at finite densities reads \cite{ISI}
\bmath
e_{xc}^{ISI}(r_s,\zeta)=\frac{a_{\infty}}{r_s} + \frac{2X}{Y}\left[
(1+Y)^{1/2}-1-Z\,\ln \left(\frac{(1+Y)^{1/2}+Z}{1+Z}\right)\right].
\label{excISI2DUG}
\emath
Using $b_{\infty}=\frac12$ and writing $e_x(r_s,\zeta)=c_x(\zeta)/r_s$, we have explicitly \cite{ISI}
\bmath
X(r_s,\zeta ) & = & \frac{-b_0(\zeta )}{[c_x(\zeta )-a_{\infty}]^2}\;\frac1{r_s},
\nonumber\\
Y(r_s,\zeta ) & = & \frac{4\,b_0(\zeta )^2}{[c_x(\zeta )-a_{\infty}]^4}\;r_s,
\nonumber\\
Z(\zeta )\;\;\;\;\; & = &
\frac{-b_0(\zeta )}{[c_x(\zeta )-a_{\infty}]^3}\; -\; 1.
\label{XYZ2D}
\emath
Eq.~\eqref{excISI2DUG} provides a simple explicit LSD,
\bmath
E^{LSD}_{xc}[\rho_\ua,\rho_\da]=\int d^2r\,\rho({\bf r})\,
                        e_{xc}^{ISI}\Big(r_s({\bf r}),\zeta({\bf r})\Big),
\label{LSD}
\emath
for treating arbitrary 2D electron systems (also finite ones such as quantum dots) by the
Kohn-Sham Equations of DFT. In Eq.~\eqref{LSD}, $r_s({\bf r})=a_B^{-1}[\pi\rho({\bf r})]^{-1/2}$
and $\zeta({\bf r})=[\rho_\ua({\bf r})-\rho_\da({\bf r})]/\rho({\bf r})$.

In Fig.~2a, the ISI prediction $e_c^{ISI}(r_s,\zeta)=e_{xc}^{ISI}-e_x$ for the correlation
energy of the unpolarized uniform electron gas ($\zeta=0$) is compared with the accurate
parametrization of the
fixed-node diffusion Monte Carlo results in Ref.~\cite{AMGB}. $e_c^{ISI}$ differs slightly from
the latter by up to 4\%. This mild deviation might be cured by including in the ISI a simple
model for the next-order coefficient of expansion \eqref{ecHD} \cite{DPI}. In the high-density
limit ($r_s\to0$), however, where the present result is exact, the parametrization in
Ref.~\cite{AMGB} has for $0.7<\zeta<0.95$ a small positive deviation \cite{AMGB}, shown in
Fig.~2b.

\newpage

\noindent
{\Large\bf Figure captions:}

\noindent
Fig.~1. (a) Numerical results (dots) for the function $f(\zeta)$ of Eq.~\eqref{fdef}
obtained by Monte Carlo integrations of expression \eqref{intMC} (with $\si_1\si_2=\ua\da$)
at selected values of $\zeta$. The analytical function $f_1(\zeta)$ of Eq.~\eqref{fzeta}
is plotted as a dashed curve. The solid curve represents the accurate fit
$f_1(\zeta)+\delta f(\zeta)$, using Eq.~\eqref{dfzeta} for $\delta f(\zeta)$.
(b) The fit \eqref{dfzeta} (solid curve) compared to the true
deviation (dots) of the Monte-Carlo-integration results from $f_1(\zeta)$.

\noindent
Fig.~2. The correlation energy of Ref.~\cite{AMGB} (dotted curves) versus the present ISI
results (solid curves).


\newpage
\pagestyle{empty}
\noindent

\begin{figure}[H]
\centering
\includegraphics[width=12cm]{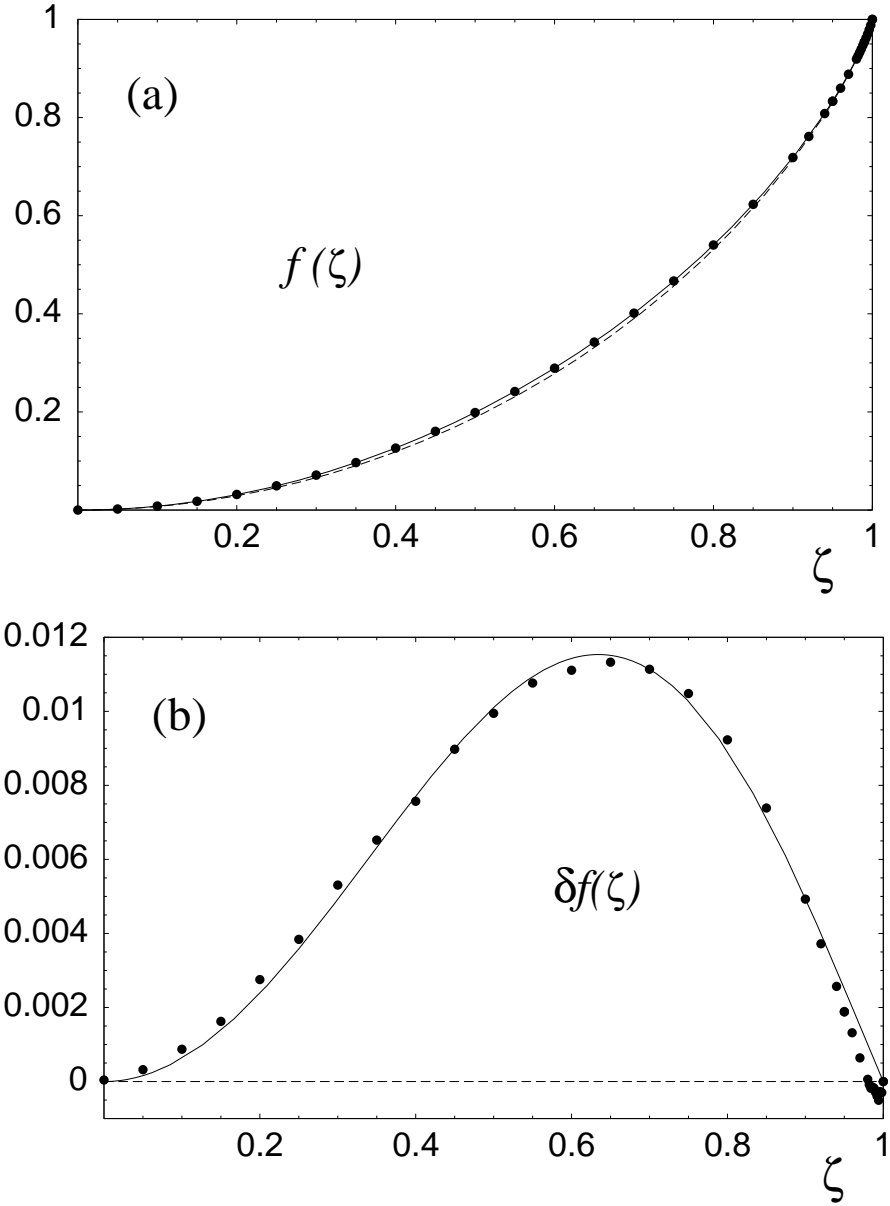}
\parbox{12.5cm}{
\caption{\label{HHcspec}
(a) Numerical results (dots) for the function $f(\zeta)$ of Eq.~\protect\eqref{fdef}
obtained by Monte Carlo integrations of expression \protect\eqref{intMC}
(with $\si_1\si_2=\ua\da$) at selected values of $\zeta$.
The analytical function $f_1(\zeta)$ of Eq.~\protect\eqref{fzeta}
is plotted as a dashed curve. The solid curve represents the accurate fit
$f_1(\zeta)+\delta f(\zeta)$, using Eq.~\protect\eqref{dfzeta} for $\delta f(\zeta)$.
(b) The fit \protect\eqref{dfzeta} (solid curve) compared to the
deviation (dots) of the Monte-Carlo-integration results from $f_1(\zeta)$.
}}
\end{figure}

\begin{figure}[H]
\centering
\includegraphics[width=12cm]{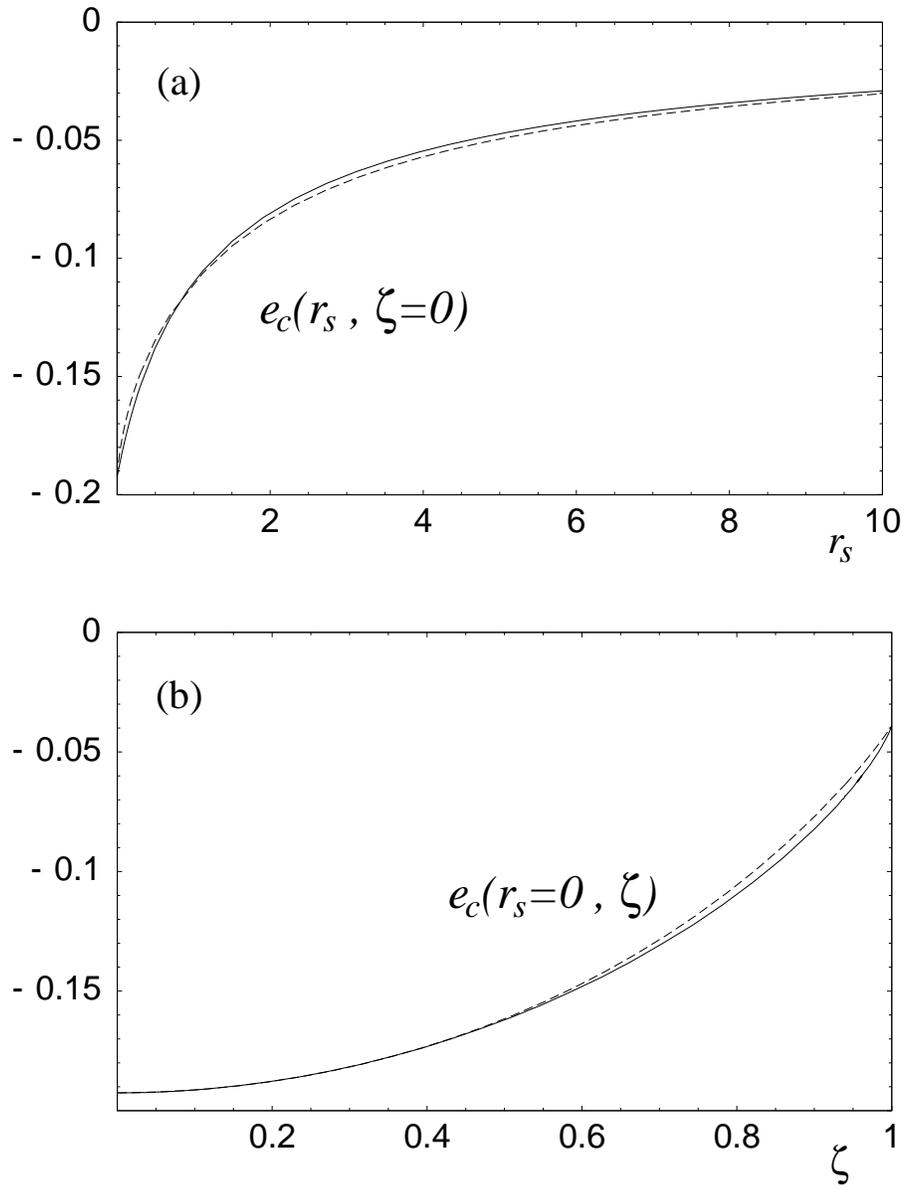}
\parbox{12.5cm}{
\caption{\label{HHcspec2}
The correlation energy of Ref.~\protect\cite{AMGB} (dotted curves) versus the present ISI
results (solid curves).}}
\end{figure}

\end{document}